\documentclass{phb-proc4-auth}
\usepackage{graphicx}
\usepackage{amssymb}
\begin{document}
\begin{frontmatter}
\journal{SCES '04}
\title{Destruction of the Kondo effect in a multi-channel
Bose-Fermi Kondo model}
\author
{Stefan Kirchner\corauthref{1},}
\author
{Lijun Zhu,}
\author
{and Qimiao Si}
\address
{Department of Physics \& Astronomy, Rice University,
 Houston, TX 77005}
\corauth[1]{Corresponding Author: kirchner@rice.edu}
\begin{abstract}
We consider the
SU($N$)$\times$SU($\kappa N$) generalization of
the 
spin-isotropic 
Bose-Fermi Kondo model 
in the limit of large $N$.
There are three fixed points corresponding to 
a multi-channel non-Fermi liquid phase,
a local spin-liquid phase, 
and 
a Kondo-destroying
quantum critical point (QCP).
We show that
the QCP has strong similarities with
its counterpart in the single-channel model,
even though the Kondo phase is very different
from the latter.
We also discuss the evolution of the dynamical
scaling properties away from the QCP.
\end{abstract}
\end{frontmatter}
Studies of heavy fermion systems within the framework of local
quantum criticality 
have largely been based on
the self-consistent 
Bose-Fermi Kondo (BFK) model \cite{Si.01}.
Here, we will consider the spin-isotropic 
multi-channel BFK model,
whose Hamiltonian is
\begin{eqnarray}
H_{\mbox{\tiny MBFK}} &=& 
(J_K/N) \sum_{\alpha}{\bf S}
\cdot {\bf s}_{\alpha}
+ \sum_{p,\alpha,\sigma} E_{p}~c_{p \alpha
\sigma}^{\dagger} c_{p \alpha \sigma}
\nonumber\\
&+& 
(g/\sqrt{N}) \sum_{p} {\bf S} \cdot
{\bf \Phi}
+ \sum_{p}
w_{p}\,{\bf \Phi}_{p}^{\;\dagger}\cdot {\bf \Phi}_{p}.
\label{H-MBFK}
\end{eqnarray}
We consider {\it fixed}
input bath spectra:
a finite $\sum_p \delta (E_F-E_p)=N_0$ for 
the conduction electrons
and a sub-ohmic spectrum for bosons,
$\sum_p\delta (\omega - w_p)
\sim \omega^{1-\epsilon}$, 
for $0< \omega < \Lambda$.
The spin and channel indices are $\sigma = 1, 
\ldots, N$
and
$\alpha=1, \ldots, M=\kappa N$,
respectively,
and ${\bf \Phi} \equiv \sum_p ( {\bf \Phi}_{p} +
{\bf \Phi}_{-p}^{\;\dagger} )$ contains $N^2-1$
components. 
The study of the large-$N$ limit was first reported
in Ref.~\cite{Zhu.04}. Our motivation is to 
access the physics of destruction of the Kondo 
effect: for the spin-isotropic
case, the only alternative method capable of doing so is 
the perturbative ($\epsilon$-expansion) renormalization
group (RG)~\cite{Zhu.02}.
The purpose of this short paper 
is two-fold. We will put the results of 
Ref.~\cite{Zhu.04} in a more general context, 
emphasizing the similarities of the quantum
critical behavior with its counterpart for the
single-channel BFK model. We will
also present the results on the dynamical
scaling away from the QCP.

As with the multi-channel Kondo 
model~\cite{Parcollet.98},
we write the local moment in the antisymmetric representation
in terms of pseudo-fermions,
$S_{\sigma \sigma'}=f_{\sigma}^{\dag}
f_{\sigma'}-\frac 12 \delta_{\sigma, \sigma'}$
and enforce the accompanying constraint, 
$\sum_{\sigma=1}^N f_{\sigma}^{\dag}f_{\sigma}= \frac 12  N$, 
by a Lagrange multiplier $i\lambda$. We then use a dynamical
Hubbard-Stratonovich field 
$B_{\alpha}(\tau)$ to represent $\sum_{\sigma}f_{\sigma}^{\dagger}
c_{\alpha \sigma} /\sqrt{N}$.
The partition function is written as $Z=Z_0Z_1$,
where $Z_0$ describes the $c$- and $\Phi$-baths 
alone; the corresponding free energy,
$F_0 \equiv -(1/\beta) \ln Z_0$, is of order $N^2$.
The effective action for $Z_1$  is 
$S _{\rm eff}
= \int d \tau 
( L_1 +  L_j + L_g)$, where
\begin{eqnarray} 
{ L}_j &=&
(1/N)
\int d\tau' 
\sum_{\alpha,\sigma} 
B_{\alpha}f_{\sigma}^{\dag}(\tau) 
{G}_{0}(\tau'-\tau)
B_{\alpha}^{\dag}f_{\sigma}(\tau') \; ,
\nonumber \\%
{ L}_g &=& 
(1/N)
\int d\tau' 
\sum_{\sigma \sigma'} 
f_{\sigma}^{\dag} f_{\sigma'}(\tau) 
{\chi}_{0}^{-1}(\tau-\tau')
f_{\sigma'}^{\dag} f_{\sigma}(\tau') 
\;, \nonumber \\%
{ L}_1 &=& 
(1/{J_K}) \sum_{\alpha} 
B_{\alpha}^{\dag}(\tau)B_{\alpha}(\tau) 
+
\sum_{\sigma} f_{\sigma}^{\dag}(\tau) 
\partial_{\tau} f_{\sigma}(\tau) 
\nonumber\\
&+& \sum_{\sigma} i\lambda [ 
f_{\sigma}^{\dag}(\tau) 
f_{\sigma}(\tau)-
1/2 ] \;,  %
\label{eq:action-multic-bfk2} 
\end{eqnarray}
where 
${ G}_0 =  - \langle T_{\tau} 
c_{\sigma\alpha}(\tau)c_{\sigma\alpha}^{\dagger}(0)\rangle _0$
and the bosonic Weiss field
$\chi_{0}^{-1} \equiv - g^2 G_{\Phi}
=-g^2
\langle T_{\tau} \Phi(\tau) \Phi^{\dagger}(0) 
\rangle _0$.
From $S_{\rm eff}$, which is of order $N$,
the dynamical saddle-point equations of
Ref.~\cite{Zhu.04} follow straightforwardly.
 \begin{figure}[t!]
\centerline{\includegraphics[width=0.55\linewidth]{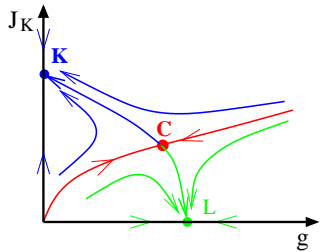}}
    \caption{
RG flows:
``K'', ``C'', and ``L'' refer to the multi-channel
Kondo, critical, and local spin-liquid fixed points, respectively.}
\label{FIG1} 
\end{figure}
In this dynamical large-N limit,
the multi-channel Kondo fixed point arises 
for the model without any bosonic bath~\cite{Parcollet.98}.
The same limit is also known to possess a local spin-liquid fixed point
for the model without conduction electrons~\cite{Vojta.00}.
Does it capture a non-trivial QCP?
The answer is {\it a priori} not clear: for instance,
the rescaling to the appropriate powers of $1/N$ 
[{\it cf.} Eq.~(\ref{H-MBFK})] could have collapsed 
the transition point to one of the axes 
in Fig.~\ref{FIG1}. For small $\epsilon$, 
we have been able to show
[using the RG method of Ref.~\cite{Zhu.02}]
that a non-trivial QCP does exist in the large-N limit.
The RG flow and the three fixed points are described 
in Fig.~\ref{FIG1}. The destruction of the (multi-channel) 
Kondo effect takes place as the separatrix is reached from
left.

The Kondo fixed point of our model describes a multi-channel
non-Fermi liquid phase, which is clearly different from the 
exactly-screened Fermi liquid phase of the 
single-channel BFK model. Nonetheless, we find that the QCPs
of the two models have strong similarities.
Consider the anomalous dimension, $\eta$, for the local
spin susceptibility [$\chi (\tau) \sim 1/{|\tau|^{\eta}}$].
At the QCP of our model, we find $\eta=\epsilon$
from both the perturbative $\epsilon$-expansion RG (to infinite orders 
in $\epsilon$) and the saddle-point analysis 
(for arbitrary $\epsilon$ in the range $0<\epsilon<1$).
This result is the same as the $\epsilon$-expansion RG result 
(also to infinite orders) for the QCP of the single-channel
BFK model. The situation is somewhat reminiscent of the
effects of spin-symmetry breaking in the single-channel
BFK model. There, the bosonic fixed point with 
Ising anisotropy (whose local susceptibility
contains a finite Curie constant~\cite{Zhu.02}) is very
different from its counterpart with SU(2) symmetry 
(whose local susceptibility is algebraic~\cite{Vojta.00}).
Yet, $\eta$ is the same for the QCP with either Ising 
anisotropy or SU(2) spin invariance~\cite{Zhu.02}.
The above results provide some justification to the usage of the 
multi-channel BFK model (which is amenable to dynamical large-$N$ 
approach) as a means to shed light on the quantum critical properties
of the single-channel
BFK model (for which no dynamical large-$N$ approach is available),
even though the latter is what is physically of relevance to
the magnetically quantum-critical heavy fermion systems.

\begin{figure}[t!]
\centerline{\includegraphics[width=1.3\linewidth]{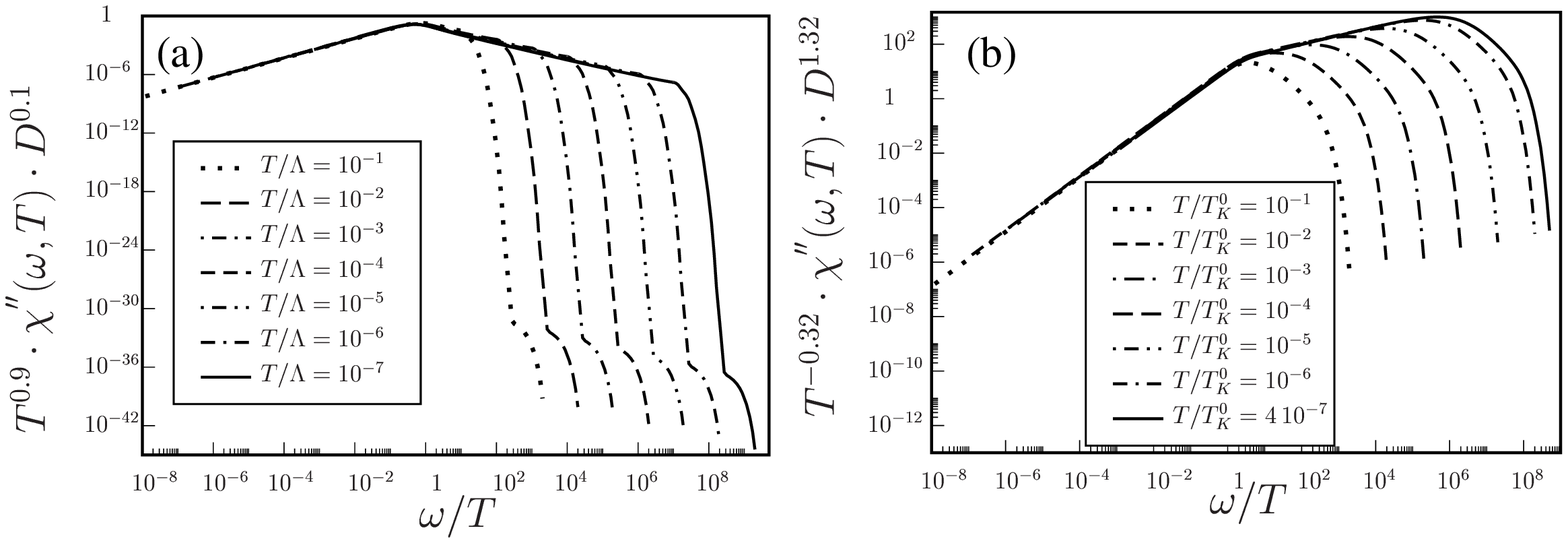}}
     \caption{The
spin susceptibility of
(a) the boson-only model, with $g=0.3$ and $\Lambda=0.05$
and (b) the fermion-only model, with $J_K=0.8$. Here $\kappa=1/2$, 
$\epsilon=0.9$, and $D \equiv 1/2N_0=1$.
} 
\label{FIG2}
 \end{figure}

An important advantage of the dynamical large-N saddle-point analysis 
lies in its ability to determine the quantum critical properties
non-perturbatively in $\epsilon$. At the QCP, the dynamical spin
susceptibility shows $\omega/T$ scaling~\cite{Zhu.04}, 
for arbitrary ratio of $\omega/T$ in almost the entire range
$\omega,T < T_K^0$, where $T_K^0$ is the bare Kondo scale. 
What happens when we move away from the QCP? To address this issue, we first
consider the behavior of the phases on both sides of the QCP.
For the bosonic (local spin-liquid) fixed point, we show in 
Fig.~\ref{FIG2}(a) the dynamical spin susceptibility for 
a finite $g$ but zero Kondo coupling. It clearly displays an $\omega/T$
scaling. Moreover, the exponent is essentially the same as that for the 
QCP. The latter implies that the local spin-susceptibility should
show $\omega/T$ scaling essentially everywhere on the right of 
the separatrix [{\it cf.} Fig.~\ref{FIG1}];
this is indeed seen in our results (not shown).

For the Kondo fixed point, we show the dynamical local
spin susceptibility
in Fig.~\ref{FIG2}(b). Here (as is already 
known~\cite{Parcollet.98})
an $\omega/T$ scaling
also obtains; this reflects the multi-channel nature of the Kondo phase
and is very different from what happens in the Kondo phase of
the single-channel model. The corresponding exponent is
different from  that for the QCP. It follows that
$\omega/T$ scaling should be violated 
in the region between the separatrix 
and the $J_K$ axis [{\it cf.} Fig.~\ref{FIG1}], as we indeed find 
in our saddle-point analysis (not shown).

We thank DFG (SK), NSF Grant No.\ DMR-0424125,
and the Robert A. Welch Foundation for support.

\end{document}